\documentstyle[preprint,fancybox,epsf,amssymb,amsbsy,amstext,amsfonts,aps]{revtex}
\draft
\tighten

\newcommand{\AmS}{{\protect\the\textfont2
  A\kern-.1667em\lower.5ex\hbox{M}\kern-.125emS}}

\def\bvare{\boldsymbol{\varepsilon}}
\def\bsig{\boldsymbol{\sigma}}
\def\sc{\scriptsize}

\begin{document}

\title{Burkhardt-Cottingham sum rule and forward spin polarizabilities 
in Heavy Baryon Chiral Perturbation Theory}

\author{Chung Wen Kao$^a$
\thanks{E-mail address:kao@theory.ph.man.ac.uk}, 
Thomas Spitzenberg$^b$  
\thanks{E-mail address: spitzenb@kph.uni-mainz.de},
and Marc Vanderhaeghen$^b$
\thanks{E-mail address: marcvdh@kph.uni-mainz.de}
}

\address{$^a$ Department of Physics and Astronomy, University of Manchester, Manchester, M13 9PL UK}
\address{$^b$ Institut f\"{u}r Kernphysik, Johannes
Gutenberg-Universit\"{a}t, D-55099 Mainz, Germany}
\date{\today}
\maketitle

\begin{abstract}
We study spin-dependent sum rules for forward virtual Compton scattering
(VVCS) off the nucleon in heavy baryon chiral perturbation theory
(HBChPT) at order ${\mathcal{O}}(p^4)$. We show how these sum rules can be
evaluated from low energy expansions (in the virtual photon energy) of
the forward VVCS amplitude. We study in particular the
Burkhardt-Cottingham sum rule in HBChPT and the higher terms in the
low energy expansion, which can be related to generalized forward spin
polarizabilities of the nucleon. 
The dependence of these observables on the photon virtuality $Q^2$ 
can be accessed, at small and intermediate $Q^2$ values, from existing
and forthcoming data at Jefferson Lab.

\end{abstract}

\section{Introduction}

Low energy (real) Compton scattering off the nucleon is a very useful
tool to investigate global properties of nucleon structure. Low energy
theorems \cite{Low54,Gel54} express the leading terms in the energy of
the forward Compton amplitude in terms of total charge, mass and
anomalous magnetic moment of the system. The higher terms in such a
low energy expansion can be identified with the response of the
nucleon to an external electromagnetic field, parametrized by dipole
and higher order nucleon polarizabilities (see Ref.~\cite{DPV02} for a
recent review and references therein). 
\newline
\indent
Sum rules for forward real Compton scattering directly connect these
low energy quantities to the nucleon excitation spectrum. Some well
known sum rules are the Baldin sum rule \cite{Bal60} which connects
the sum of electric and magnetic nucleon polarizabilities to the total
photoabsorption cross section on the nucleon, as well as the
Gerasimov-Drell-Hearn (GDH) sum rule \cite{Ger65,Dre66}. 
The latter sum rule establishes a connection
between the nucleon anomalous magnetic moment and the helicity
difference cross section for photon scattering with helicity parallel
or anti-parallel to the nucleon helicity. The GDH sum rule has been the
subject of several experiments in recent years \cite{Ahr01,Hel02}.
\newline
\indent
It has also been emphasized that the above sum rules can be 
generalized to the virtual photon case, see e.g. Refs.~\cite{Ede98,Ji01,DPV02}.
For forward scattering of spacelike virtual photons (with virtuality
$Q^2$) on the nucleon, the corresponding sum rules relate nucleon 
structure quantities to inclusive electroproduction cross sections. 
At large $Q^2$, they yield the sum rules studied in deep-inelastic
scattering (DIS) experiments (see Ref.~\cite{FJ01} for a recent
review), such as the Bjorken sum rule \cite{Bj66} or the
Burkhardt-Cottingham (BC) sum rule \cite{BC70}. 
The Bjorken sum rule involves the difference of the first moments 
of the proton and neutron helicity structure functions $g_1^p - g_1^n$, 
and has been verified by several experiments \cite{FJ01}.
The BC sum rule on the other hand, which involves the first moment of
the second spin dependent structure function $g_2$ of the nucleon, has
been addressed by a dedicated experiment only very recently \cite{E155X}. 
So far, the experiment does not allow for a conclusive test of the BC sum
rules due to an unmeasured small-$x$ region and awaits further experiments. 
\newline
\indent
Having measured forward Compton sum rules at the real photon point and
at large $Q^2$, through DIS, one may ask the question how these sum
rules interpolate between both limits when one varies $Q^2$, which
plays the role of the spatial resolution at which one probes the
nucleon. 
The study of such sum rules for the virtual photon nucleon Compton
amplitudes as function of $Q^2$ from the real photon point to large
$Q^2$, opens up the perspective to map out in detail the transition
from the resonance dominated regime at low $Q^2$ to the partonic
regime at large $Q^2$, described by perturbative QCD. The theoretical
tool to analyse the large $Q^2$ regime is given by the operator
product expansion (OPE). At low $Q^2$, below about 0.5 GeV$^2$, where one is
in the non-perturbative realm of QCD, a theoretical tool to
investigate these sum rules is provided by chiral perturbation theory
(ChPT). Using ChPT, the first moment of $g_1$ has been studied in
Refs.~\cite{Ji01,JKO00,BHM02}. 
It has been pointed out in Ref.~\cite{Bur01} that in the first moment of
the proton - neutron difference $g_1^p - g_1^n$ the $\Delta$(1232) and
other isospin 3/2 resonances drop out, and therefore this 
might be a promising observable to look for a smooth transiton between 
the ChPT result at low $Q^2$ and OPE description at large $Q^2$.  
\newline
\indent
In this letter, we study in the 
framework of the heavy-baryon ChPT (HBChPT) 
to order ${\mathcal{O}}(p^4)$, the first
moment of the structure function $g_2$ and the BC sum rule, as well as
sum rules involving higher moments of the spin dependent structure
functions, which have the physical interpretation of 
generalized forward spin polarizabilities. 
\newline
\indent
In Section~2, we start with the general formalism for spin
dependent forward virtual Compton scattering (VVCS). For the two
spin-dependent VVCS amplitudes, we write down dispersion relations (DRs) and
discuss the nucleon pole contribution. For the non-pole contributions,
we show that one can proceed to write down a low energy expansion
(LEX) in the virtual photon energy. The DRs allow us to write down
sum rules for each term in such a LEX. The lowest terms in the LEX are
given by the first moments of the nucleon structure functions $g_1$
and $g_2$, and we show how sum rules for the higher terms can be
interpreted as generalized forward spin polarizabilities. 
\newline
\indent
In Sections~3 and 4, we then proceed and calculate the terms in such a
LEX in HBChPT. In particular in Section~3, we calculate the $Q^2$
dependence of the lowest moment of the nucleon structure function
$g_2$ and discuss the validity of the BC sum rule in HBChPT at order 
${\mathcal{O}}(p^4)$. 
\newline
\indent
In Section~4, we present the results for the higher terms in the LEX
in  HBChPT at order ${\mathcal{O}}(p^4)$, and give the corresponding
expressions for the generalized forward spin polarizabilities of the
nucleon. We study the convergence of the HBChPT expansion by comparing 
the results at order ${\mathcal{O}}(p^3)$ and ${\mathcal{O}}(p^4)$.
We also compare the $Q^2$ dependence of the HBChPT results with those
of a phenomenological resonance estimate, using the MAID model \cite{MAID}.
\newline
\indent
Finally, in Section~5 we present our conclusions.

\section{Spin dependent forward virtual Compton scattering (VVCS)}

We start with the spin-dependent doubly virtual Compton scattering
amplitude in the forward direction (VVCS) as given in \cite{DPV02}~:
\begin{equation}
\label{eq:vvcsgttglt}
T(\nu,\,Q^2,\,\theta=0)^{spin} \,=\, 
\;i{\bsig}\cdot({\bvare}\,'^{\ast}\times{\bvare}) \;g_{TT}(\nu,\,Q^2)
\;-\; i{\bsig}\cdot[({\bvare}\,'^{\ast}-{\bvare})\times \hat{q} \,]
\; g_{LT}(\nu,\,Q^2) \, ,
\end{equation}
where $\nu$ is the virtual photon energy, $Q^2$ its virtuality, and 
$\hat{q}$ the unit vector along the virtual photon momentum. 
Furthermore, $\bvare$ ($\bvare \,'$) are the polarization vectors of
the initial (final) photons. 
In the VVCS amplitudes $g_{TT}$ and $g_{LT}$, T (L) denote the transverse
(longitudinal) virtual photon polarizations. In order to relate with
the usual nucleon structure functions $g_1$ and $g_2$, it is also 
useful to cast Eq.~(\ref{eq:vvcsgttglt}) in the following covariant form~:
\begin{eqnarray}
\label{eq:vvcss1s2}
T(\nu,\,Q^2,\,\theta=0)^{spin} \,=\,
\varepsilon_{\mu}'^{\ast} \; \varepsilon_{\nu} 
&&\left \{ \frac{i}{M}\,\epsilon^{\mu\nu\alpha\beta}\,q_{\alpha}
s_{\beta}\, S_1(\nu,\,Q^2) \right . \nonumber \\
&&\left . + \frac{1}{M^3}\,\epsilon^{\mu\nu\alpha\beta}\,q_{\alpha}
(p\cdot q\ s_{\beta}-s\cdot q\ p_{\beta})\, S_2(\nu,\,Q^2) \right
\}\, ,
\end{eqnarray}
where $s^\alpha$ is the nucleon covariant spin vector satisfying
 $s^2$ = -1 and $s \cdot p$ = 0, with $p$ ($q$) the nucleon (photon) 
four-momenta. 
The relation between the spin-dependent amplitudes $g_{TT}$, $g_{LT}$
 and $S_1, S_2$ is given by \cite{DPV02}~:
\begin{eqnarray}
\label{eq:s1gttglt}
S_1(\nu,\,Q^2) & = &
\frac{\nu\,M}{\nu^2+Q^2}\left(g_{TT}(\nu,\,Q^2) +
\frac{Q}{\nu}\,g_{LT}(\nu,\,Q^2)\right) \, , \\
\label{eq:s2gttglt}
S_2(\nu,\,Q^2) & = & -
\frac{M^2}{\nu^2+Q^2}\left(g_{TT}(\nu,\,Q^2) -
\frac{\nu}{Q}\,g_{LT}(\nu,\,Q^2)\right) \, .
\end{eqnarray}
\newline
\indent
Next, we can write down dispersion relations (DR) for the VVCS
amplitudes $g_{TT}$ and $g_{LT}$. Assuming that 
$g_{TT}(\nu, Q^2)$ and $g_{LT}(\nu, Q^2)$ drop sufficiently fast for 
$\nu \to \infty$ to ensure the convergence of the integrals (see
Ref.~\cite{DPV02} for a detailed discussion), one can
write down unsubtracted DRs for $g_{TT}$ and $g_{LT}$, which take the
form~:
\begin{eqnarray}
\label{eq:drgtt}
{\mbox{Re}}\ g_{TT}\,(\nu,\,Q^2) \,&=&\,
{\mbox{Re}}\ g_{TT}^{\mbox{\sc{pole}}}(\nu,\,Q^2)
\;+\;\frac{\nu}{2\pi^2}\,{\mathcal{P}}\,
\int_{\nu_0}^{\infty}\,\frac{K(\nu',\,Q^2)\,\sigma_{TT}(\nu',\,Q^2)}
{\nu'^2-\nu^2} \, d\nu' \, , \\
\label{eq:drglt}
{\mbox{Re}}\ g_{LT}(\nu,\,Q^2) \,&=&\,
{\mbox{Re}}\ g_{LT}^{\mbox{\sc{pole}}}(\nu,\,Q^2) \;+\;
\frac{1}{2\pi^2}\,{\mathcal{P}}\,\int_{\nu_0}^{\infty}\,
\frac{\nu'K\,(\nu',\,Q^2)\,\sigma_{LT}(\nu',\,Q^2)}{(\nu'^2-\nu^2)}\,d\nu'
\, ,
\end{eqnarray}
where the first terms in Eqs.~(\ref{eq:drgtt},\ref{eq:drglt}) are 
the pole or elastic contributions due to the $s$- and $u$-channel
singularities at $\nu = \pm \nu_B$, with $\nu_B \equiv Q^2 / (2 M_N)$, 
where $M_N$ is the nucleon mass.
These pole contributions are given by~:
\begin{eqnarray}
\label{eq:gttpole}
{\mbox{Re}}\ g_{TT}^{\mbox{\sc{pole}}} (\nu,\,Q^2) & = &
-\frac{\alpha_{\mbox{\sc{em}}} \, \nu}{2 \, M_N^2}
\frac{Q^2}{\nu^2-\nu_B^2} \,G_M^2(Q^2) \, , \\
\label{eq:gltpole}
{\mbox{Re}}\ g_{LT}^{\mbox{\sc{pole}}} (\nu,\,Q^2) & = &
\, - \, \frac{\alpha_{\mbox{\sc{em}}} \, Q}{2 \, M_N^2}
\frac{Q^2}{\nu^2-\nu_B^2} \,G_E(Q^2) \, G_M(Q^2) \, ,
\end{eqnarray}
with $\alpha_{\mbox{\sc{em}}}$ the fine structure constant
( $\alpha_{\mbox{\sc{em}}} = 1/137$ ), and 
$G_E (G_M)$ the nucleon electric (magnetic) form factors.
The dispersion integrals in Eqs.~(\ref{eq:drgtt},\ref{eq:drglt})
correspond to the cut contributions along the real axis and start at
the threshold for pion production
$\nu_0=m_{\pi}+(m_{\pi}^2+Q^2)/2M_N$, with $m_\pi$ the pion mass.
The integrals in Eqs.~(\ref{eq:drgtt},\ref{eq:drglt})
involve the product of the partial cross sections 
$\sigma_{TT}$ and $\sigma_{LT}$, mulitiplied by a photon flux factor
$K$ (with dimension of energy)
\footnote{Note that the partial cross sections $\sigma_{TT}$ and 
$\sigma_{LT}$ depend on the virtual photon flux convention. However, 
in the dispersion integrals only 
the products $K \cdot \sigma_{TT}$ and $K \cdot \sigma_{LT}$ enter,  
which are independent of this convention.}.
These partial cross sections are related to the nucleon structure
functions $g_1$ and $g_2$ as~:
\begin{eqnarray}
K \cdot \sigma_{TT} \,&=&\, \frac{4\pi^{2}\alpha_{em}}{M_N}  
\biggl( g_{1}(x, Q^2) -\gamma^{2}g_{2}(x, Q^2) \biggr), 
\label{eq:stt} \\
K \cdot \sigma_{LT} \,&=&\, \frac{4\pi^{2}\alpha_{em}}{M_N} 
\gamma \biggl( g_{1}(x, Q^2) + g_{2}(x, Q^2) \biggr) \, ,
\label{eq:slt}
\end{eqnarray}
with $\gamma \equiv Q/\nu$ and $x \equiv Q^2 / (2 M_N \, \nu)$.
\newline
\indent
For the non-pole contributions to $g_{TT}$ and $g_{LT}$, one can
perform a low energy expansion (LEX) as follows \cite{DPV02}~:
\begin{eqnarray}
{\mbox{Re}}\ g_{TT}(\nu,\,Q^2) \,-\,
{\mbox{Re}}\ g_{TT}^{\mbox{\sc{pole}}}(\nu,\,Q^2) &=&
\left(\frac{2 \, \alpha_{\mbox{\sc{em}}} }{M_N^2} \right) \, I_A(Q^2) \; \nu
\,+\, \gamma_0(Q^2) \; \nu^3 \,+\, {\mathcal{O}}(\nu^5) \, , 
\label{eq:gttlex} \\
{\mbox{Re}}\ g_{LT}(\nu,\,Q^2) -
{\mbox{Re}}\ g_{LT}^{\mbox{\sc{pole}}}(\nu,\,Q^2) &=&
\left(\frac{2 \, \alpha_{\mbox{\sc{em}}} }{M_N^2} \right) Q \, I_3(Q^2)
\,+\, Q \, \delta_{LT}(Q^2) \, \nu^2 \,+\, {\mathcal{O}}(\nu^4) \, .
\label{eq:gltlex}
\end{eqnarray}
For the ${\mathcal{O}}(\nu)$ term in Eq.~(\ref{eq:gttlex}), one
obtains from Eq.~(\ref{eq:drgtt}) a generalization of the GDH sum rule
as~:
\begin{eqnarray}
\label{eq:}
I_A(Q^2) &\,=\,&
\frac{M_N^2}{4 \, \pi^2 \, \alpha_{\mbox{\sc{em}}}}\,
\int_{\nu_0}^{\infty}\, \frac{K(\nu, \, Q^2)}{\nu} \,
\frac{\sigma_{TT}\,(\nu,\,Q^2)}{\nu}\,d\nu \, , \nonumber\\
&\,=\,& \frac{2 \, M_N^2}{Q^2}\,
\int_{0}^{x_0}\,dx \, \left\{ g_1\,(x,\,Q^2)
\,-\, \frac{4 M_N^2}{Q^2} \, x^2 \, g_2\,(x,\,Q^2) \right\} \, ,
\end{eqnarray}
and recovers the GDH sum rule at $Q^2$ = 0, as  $I_A(0) = - \kappa_N^2
/ 4$, with $\kappa_N$ the nucleon anomalous magnetic moment ($\kappa_p
= 1.79$, $\kappa_n = -1.91$).
For the ${\mathcal{O}}(\nu^0)$ term in Eq.~(\ref{eq:gltlex}), one
obtains from Eq.~(\ref{eq:drglt}) the sum rule
\begin{eqnarray}
\label{eq:i3int}
I_3(Q^2) \,&=&\,
\frac{M_N^2}{4 \, \pi^2 \, \alpha_{\mbox{\sc{em}}} }\,
\int_{\nu_0}^{\infty}\,\frac{K(\nu \, , Q^2)}{\nu} \,
\frac{1}{Q} \, \sigma_{LT}\,(\nu,\,Q^2)\,d\nu\  \, , \nonumber\\
\,&=&\, \frac{2 \, M_N^2}{Q^2}\,\int_{0}^{x_0}\,dx \,
\left\{ g_1\,(x,\,Q^2) \,+\, g_2\,(x,\,Q^2) \right\} \, .
\end{eqnarray}
For the sum rule of Eq.~(\ref{eq:i3int}) to exist, one sees that 
$\sigma_{LT}$ should vanish faster than $1 / \nu$ at large $\nu$.
\newline
\indent
The higher order terms in Eqs.~(\ref{eq:gttlex}) and (\ref{eq:gltlex}) 
can be expressed in terms of nucleon spin polarizabilities as \cite{DPV02}~:
\begin{eqnarray}
\label{eq:gammao}
\gamma_0\,(Q^2) \,&=&\, \frac{1}{2\pi^2}\,
\int_{\nu_0}^{\infty}\, \frac{K(\nu, \, Q^2)}{\nu} \,
\frac{\sigma_{TT}\,(\nu,\,Q^2)}{\nu^3}\,d\nu \, , \nonumber \\
\,&=&\, \frac{\alpha_{\mbox{\sc{em}}} \, 16 \, M_N^2}{Q^6}\,
\int_{0}^{x_0}\,dx \, x^2 \, \left\{ g_1\,(x,\,Q^2)
\,-\, \frac{4 M_N^2}{Q^2} \, x^2 \, g_2\,(x,\,Q^2) \right\} \, ,
\end{eqnarray}
and 
\begin{eqnarray}
\label{eq:deltalt}
\delta_{LT}\,(Q^2) \,&=&\, \frac{1}{2\pi^2}\,
\int_{\nu_0}^{\infty}\,\frac{K(\nu, \, Q^2)}{\nu} \,
\frac{\sigma_{LT}(\nu\,Q^2)}{Q\,\nu^2}\,d\nu  \, , \nonumber \\
\,&=&\, \frac{\alpha_{\mbox{\sc{em}}} \, 16 \, M_N^2}{Q^6}\,
\int_{0}^{x_0}\,dx \, x^2 \,
\left\{ g_1\,(x,\,Q^2) \,+\, g_2\,(x,\,Q^2) \right\} \, .
\end{eqnarray}
\newline
\indent
Analogously, we can also write down corresponding LEX and sum rules
for the spin dependent Compton amplitudes $S_1$ and $S_2$ of 
Eq.~(\ref{eq:vvcss1s2}). For these, one again first has to separate 
the pole contributions to $S_1$ and $\nu S_2$, which are given by
\begin{eqnarray}
{\rm Re}\ S_1^{\mbox{\sc{pole}}} (\nu,\,Q^2) \,&=&\,
-\frac{\alpha_{\mbox{\sc{em}}}}{2M_N}
\frac{Q^2}{\nu^2-\nu_B^2}\,F_D(Q^2) \, \left(F_D(Q^2) + F_P(Q^2) \right ) \, ,
\\
{\rm Re}\ \left( \nu S_2 (\nu,\,Q^2) \right)^{\mbox{\sc{pole}}} \,&=&\,
\frac{\alpha_{\mbox{\sc{em}}}}{2}
\frac{\nu_B^2}{\nu^2-\nu_B^2}\,F_P(Q^2) \, 
\left(F_D(Q^2) + F_P(Q^2) \right ) \, ,
\end{eqnarray}
with $F_D$ ($F_P$) the nucleon Dirac (Pauli) form factors.
The resulting LEX for the non-pole contributions are then 
given by \cite{DPV02}~:
\begin{eqnarray}
&&\hspace{-.25cm}{\mbox{Re}}\ S_1(\nu,\,Q^2) -
{\mbox{Re}}\ S_1^{\mbox{\sc{pole}}}(\nu,\,Q^2) \,=\, \nonumber \\
&&\hspace{-.25cm}
\left(\frac{2 \, \alpha_{\mbox{\sc{em}}} }{M_N} \right) \, I_1(Q^2)
\,+\, \left[ \left(\frac{2 \, \alpha_{\mbox{\sc{em}}} }{M_N} \right) 
{1 \over {Q^2}} \left( I_A(Q^2) - I_1(Q^2) \right)
\,+\, M_N \delta_{LT}(Q^2) \right] \, \nu^2  \,+\, {\mathcal{O}}(\nu^4) ,
\label{eq:s1lex}
\end{eqnarray}
and 
\footnote{Note that the relation of Ref.~\cite{DKT01}, i.e., 
$I_A'(0) - I_1'(0) = M_N^2 / (2 \, \alpha_{\mbox{\sc{em}}}) \, 
\cdot \left( \gamma_0(0) - \delta_{LT}(0) \right)$ ensures that the 
$\nu^4$ term in $\nu S_2$ has no singularity at $Q^2 = 0$.}~:
\begin{eqnarray}
&&{\mbox{Re}}\ \nu \, S_2(\nu,\,Q^2) -
{\mbox{Re}}\ (\nu \, S_2(\nu,\,Q^2))^{\mbox{\sc{pole}}} \,=\,\nonumber \\
&& \left(2 \, \alpha_{\mbox{\sc{em}}} \right) \, I_2(Q^2)
\,-\, \left(2 \, \alpha_{\mbox{\sc{em}}} \right) 
{1 \over {Q^2}} \left( I_A(Q^2) - I_1(Q^2) \right) \, \nu^2  
\nonumber \\
&&+\, {1 \over Q^2} \left[ \left(2 \, \alpha_{\mbox{\sc{em}}} \right) 
{1 \over {Q^2}} \left( I_A(Q^2) - I_1(Q^2) \right)
\,+\, M_N^2\,\left( \delta_{LT}(Q^2) - \gamma_0(Q^2) \right) \right] \, \nu^4  
\,+\, {\mathcal{O}}(\nu^6) \, .
\label{eq:s2lex}
\end{eqnarray}
The lowest order term in Eq.~(\ref{eq:s1lex}) is connected to the
first moment of $g_1$ through the sum rule~: 
\begin{eqnarray}
I_1(Q^2) \,&\equiv&\, \frac{2M_N^2}{Q^2}\int_0^{x_0}
g_1(x,\,Q^2)\,dx \, , \nonumber \\
\,&=&\, \frac{M_N^2}{4 \, \pi^2 \, \alpha_{\mbox{\sc{em}}} }\,
\int_{\nu_0}^{\infty}\,\frac{K(\nu, Q^2)}{(\nu^2 + Q^2)}
\left\{\sigma_{TT}\,(\nu,\,Q^2) \,+\,
\frac{Q}{\nu} \, \sigma_{LT}\,(\nu,\,Q^2) \right\} \,d\nu \, ,
\label{eq:I1} 
\end{eqnarray}
where the integral over $x$ runs 
up to the value $x_0$ corresponding to $\nu_0$. 
At $Q^2 = 0$, Eq.~(\ref{eq:I1}) also reduces to the GDH sum rule,
i.e., $I_1(0) = - \kappa_N^2 / 4$.
Correspondingly, the lowest order term in Eq.~(\ref{eq:s2lex}) is
connected to the first moment of $g_2$ through the sum rule~:
\begin{eqnarray}
I_2(Q^2) \,&\equiv&\, \frac{2 \,M_N^2}{Q^2} \, 
\int_0^{x_0}g_2(x,\,Q^2)\,dx \, , 
\nonumber \\
\,&=&\, \frac{M_N^2}{4 \, \pi^2 \, \alpha_{\mbox{\sc{em}}} }\,
\int_{\nu_0}^{\infty}\,\frac{K(\nu,\,Q^2)}{\nu^2 + Q^2}
\, \left\{ \,- \sigma_{TT}(\nu,\,Q^2) \,+\,
\frac{\nu}{Q} \, \sigma_{LT}(\nu,\,Q^2) \, \right\} \, d \nu \, . 
\label{eq:I2} 
\end{eqnarray}
Combining Eqs.~(\ref{eq:I1}) and (\ref{eq:I2}), one obtains 
Eq.~(\ref{eq:i3int}), i.e., 
\begin{eqnarray}
I_3(Q^2) \,=\, I_1(Q^2) \,+\, I_2(Q^2) \, .
\end{eqnarray}
Furthermore, one sees that the higher order terms 
in Eqs.~(\ref{eq:s1lex}) and (\ref{eq:s2lex})
are expressed in terms of the integrals 
$I_1, I_A$ and the spin polarizabilities 
$\gamma_0$ and $\delta_{LT}$ introduced before. 
\newline
\indent
For the first moment of $I_2$ in Eq.~(\ref{eq:I2}) to converge, one
has to assume, as in the case of $I_3$, the strong convergence
condition that $\sigma_{LT}$ vanishes faster than $1/\nu$ at large
$\nu$. If this assumption holds, then $I_2$ satisfies a sum rule,
derived by Burkhardt and Cottingham \cite{BC70} (BC sum rule), which
allows one to express it at any $Q^2$ in terms of elastic nucleon
form factors as~:
\begin{eqnarray}
\label{eq:bc}
I_2(Q^2) \;=\;
\frac{1}{4} \, F_P(Q^2) \, \left( F_D(Q^2) + F_P(Q^2) \right) \, .
\end{eqnarray}
The BC sum rule has been shown to be satisfied in the case of quantum
electrodynamics by a calculation in lowest order of 
$\alpha_{\mbox{\sc{em}}}$ \cite{Tsai75}.
In perturbative QCD, the BC sum rule was calculated for a quark
target to first order in the strong coupling and also shown to hold 
\cite{Alt94}. 
In the next section, we investigate next the BC sum rule at small momentum
transfer within the framework of HBChPT to ${\mathcal{O}}(p^4)$.

\section{BC sum rule in HBChPT to ${\mathcal{O}}(p^4)$}

\begin{figure}[h]
\epsfysize=13cm
\centerline{\epsffile{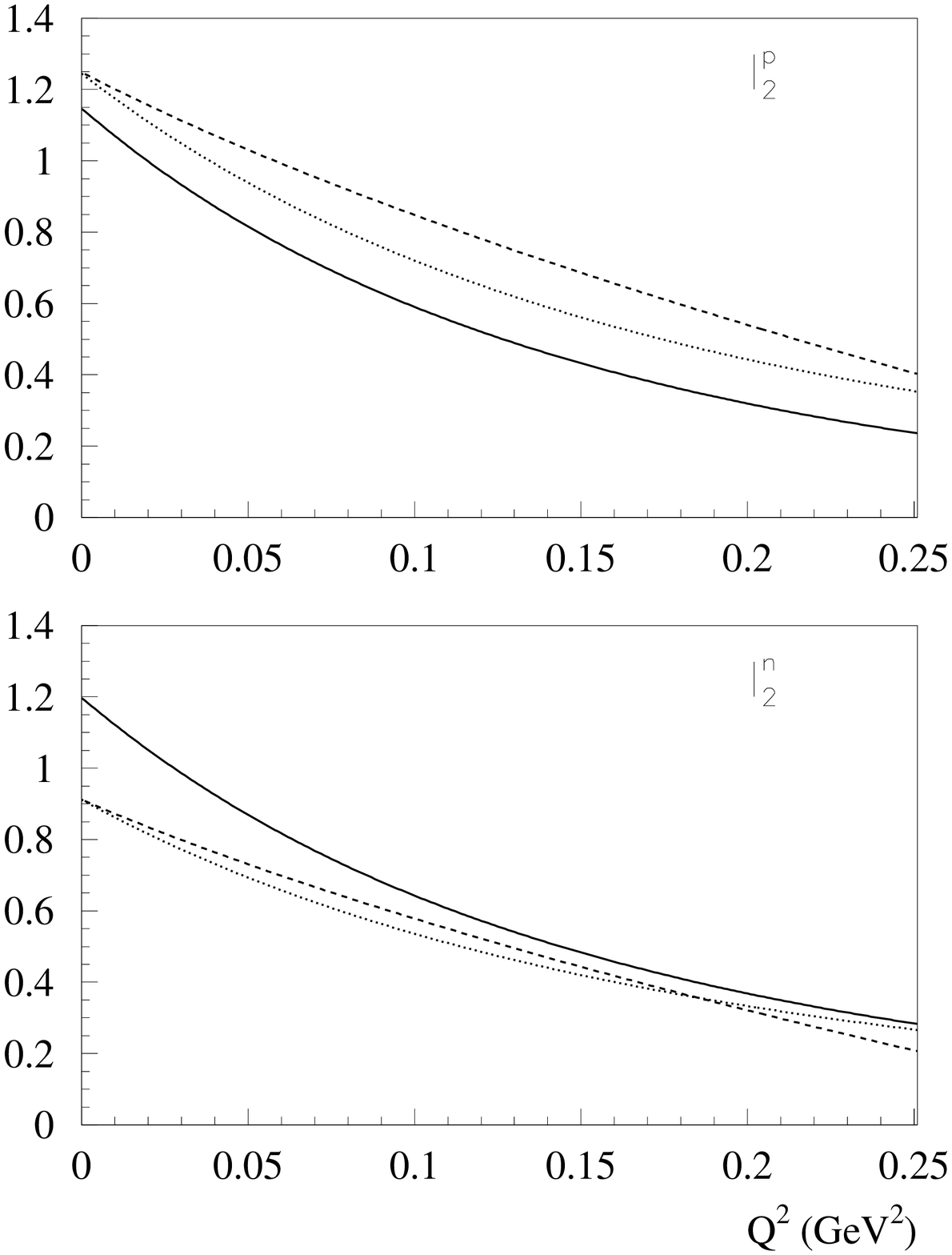}}
\vspace{0cm}
\caption{$Q^2$ dependence of the integral $I_2$ 
for proton (upper panel) and neutron (lower panel).
The solid curves represent the MAID estimate~\protect\cite{MAID,DKT01}
for the $\pi$ channel.
The dashed curves are the ${\mathcal O}(p^3) + {\mathcal O}(p^4)$ 
HBChPT results; and the dotted curves are the BC sum rule prediction
for $I_2$, evaluated with phenomenological form factors. }  
\label{fig:i2}
\end{figure}

The validity of the BC sum rule can be tested in HBChPT, 
by calculating both sides of Eq.~(\ref{eq:bc}). 
To this end, the complete one-loop calculation of the forward 
VVCS amplitudes at low energy $\nu$ and small momentum transfer $Q^2$
has been done, and some results have been presented in
Refs.~\cite{Ji01,JKO00}. In particular, the first term in the
expression of Eq.~(\ref{eq:s1lex}) for the inelastic part of 
$S_1$ has been found as~\cite{Ji01,JKO00} 
\begin{eqnarray}
I_{1}(Q^2)
&=&- \frac{1}{16}[(\kappa_{s} + \kappa_{v}\tau_{3})^2] \nonumber \\
&+&\frac{g_{A}^2 m_{\pi}M_{N}}{(4\pi F_{\pi})^2}\cdot\frac{\pi}{32}
\{(-10-12\kappa_{v})+(-2-12\kappa_{s})\tau_{3} \nonumber \\
&&\hspace{2.5cm}+\,[(20+24\kappa_{v})+(4+24\kappa_{s})\tau_{3}]
\cdot\frac{1}{w}\tan^{-1}[\frac{w}{2}] \nonumber \\ 
&&\hspace{2.5cm}+\,[(3+6\kappa_{v})+(3+10\kappa_{s})\tau_{3}]\cdot
w\tan^{-1}[\frac{w}{2}]\} \, , 
\label{eq:i1chpt}
\end{eqnarray}
with $w=\sqrt{\frac{Q^2}{m_{\pi}^{2}}}$.
Eq.~(\ref{eq:i1chpt}) is expressed 
in terms of the renormalized isoscalar (isovector) anomalous magnetic
moments $\kappa_s$ ($\kappa_v$), whose physical values are given by 
$\kappa_s = -0.12$ and $\kappa_v = 3.70$.
Furthermore, throughout this paper we use the values~:
$g_A = 1.267$, $F_\pi = 0.0924$~GeV, and $m_\pi = 0.14$~GeV. 
\newline
\indent
For the inelastic part of the amplitude $(\nu S_2)$, we found the
first term of Eq.~(\ref{eq:s2lex}) to 
next-to-leading order in HBChPT to be given by~:
\begin{eqnarray}
I_{2}(Q^2)
&=&\frac{1}{16}[(\kappa_{s}+\kappa_{v}\tau_{3})(1+\tau_{3})+(\kappa_{s}+\kappa_{v}\tau_{3})^{2}] \nonumber \\
&-&\frac{g_{A}^2 m_{\pi}M_{N}}{(4\pi F_{\pi})^2}\cdot\frac{\pi}{16}
\{(-2-4\kappa_{v})+(-2-4\kappa_{s})\tau_{3} \nonumber \\
&&\hspace{2.5cm}+ \,[(4+8\kappa_{v})+(4+8\kappa_{s})\tau_{3}]
\cdot\frac{1}{w}\tan^{-1}[\frac{w}{2}] \nonumber \\
&&\hspace{2.5cm}+ \,[(1+2\kappa_{v})+(1+2\kappa_{s})\tau_{3}]\cdot
w\tan^{-1}[\frac{w}{2}]\} \, .
\label{eq:i2chpt}
\end{eqnarray}
This result is solely from the one-particle-reducible diagrams.
To check the BC sum rule of Eq.~(\ref{eq:bc}), we need the
corresponding expressions for the form factors in HBChPT which, to the
order needed, are given by~\cite{BKM95}
\begin{eqnarray}
F_{D}[Q^2]&=&\frac{1}{2}\left[1+\tau_{3}\right], \\
F_{P}[Q^2]&=&\frac{1}{2}\left[\kappa_{s}+\kappa_{v}\tau_{3}\right]-\tau_{3}\frac{g_{A}^{2}m_{\pi}M_{N}}{(4\pi F_{\pi})^2}\cdot 2\pi\left[\frac{-1}{2}
+(\frac{w}{4}+\frac{1}{w})\tan^{-1}[\frac{w}{2}]\right]. 
\end{eqnarray}
One therefore obtains for the {\it rhs} of  Eq.~(\ref{eq:bc}) the
expression~:
\begin{eqnarray}
\label{eq:bctest}
{1 \over 4}\, F_{P}\cdot(F_{D}+F_{P})&=&\frac{1}{16}\left[\kappa_{s}(1+\kappa_{s})+\kappa_{v}(1+\kappa_{v})+[
\kappa_{s}+\kappa_{v}+2\kappa_{s}\kappa_{v}]\tau_{3}\right] \nonumber \\
&-&[(1+2\kappa_{v})+(1+2\kappa_{s})\tau_{3}]\frac{g_{A}^{2}m_{\pi}M_{N}}{(4\pi
F_{\pi})^2}\cdot {\pi \over 4} \left[\frac{-1}{2}
+(\frac{w}{4}+\frac{1}{w})\tan^{-1}[\frac{w}{2}]\right]\, . \nonumber\\  
\end{eqnarray}
By comparing Eq.~(\ref{eq:bctest}) with Eq.~(\ref{eq:i2chpt}), we
verify the BC sum rule explicitly up to NLO in HBChPT, 
i.e., to ${\mathcal{O}}(p^4)$.
\newline
\indent
The HBChPT results for the integral $I_2$ for proton and neutron 
are shown in Fig.~\ref{fig:i2}. They are compared with the results of
a resonance estimate, and with the BC sum rule value 
({\it rhs} of Eq.~(\ref{eq:bc})), evaluated with the phenomenological
proton and neutron form factors. The resonance estimate is performed
by calculating the absorption integrals on the {\it rhs} of 
Eq.~(\ref{eq:I2}), up to a maximum total {\it c.m.} energy  
$W = 2$~GeV, using the MAID model \cite{MAID} for the one-pion production
cross sections. The one-pion production channels are expected to
dominate the integral in the low $Q^2$ region. 
From Fig.~\ref{fig:i2}, one sees that when going to larger $Q^2$, 
the HBChPT result remains numerically close to the BC sum rule
value up to $Q^2 \simeq 0.25$~GeV$^2$. Moreover, also the
phenomenological MAID estimate for $I_2$ of the proton remains within 
10~\% of the BC sum rule value. For the neutron, the phenomenological
estimate displays a larger deviation in particular at the real photon
point, where it lies about 30~\% above the BC sum rule value. This may
originate from the poorly known contribution of $\sigma_{LT} / Q$ in 
the integrand of Eq.~(\ref{eq:I2}), which survives in the limit $Q^2
\to 0$.

\section{Generalized forward spin polarizabilities 
in HBChPT to ${\mathcal{O}}(p^4)$}

\begin{figure}[h]
\epsfxsize=15cm   
\centerline{\epsffile{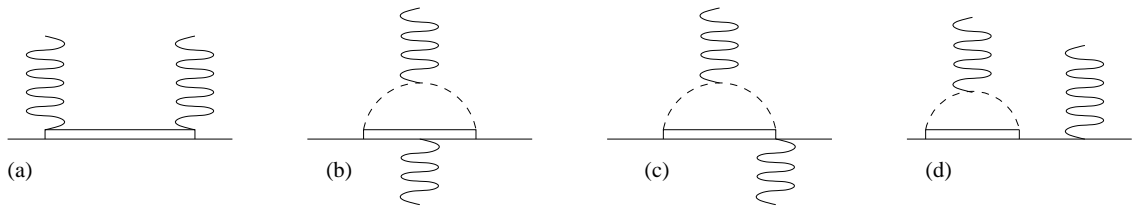}}
\vspace{0.5cm} 
\caption{The ${\mathcal O}(\varepsilon^{3})$ diagrams of
photon-nucleon Compton scattering with the $\Delta$ as intermediate
state. The crossed diagrams are not shown.} 
\label{fig:delta}
\end{figure}

\begin{figure}[h]
\epsfysize=13cm
\centerline{\epsffile{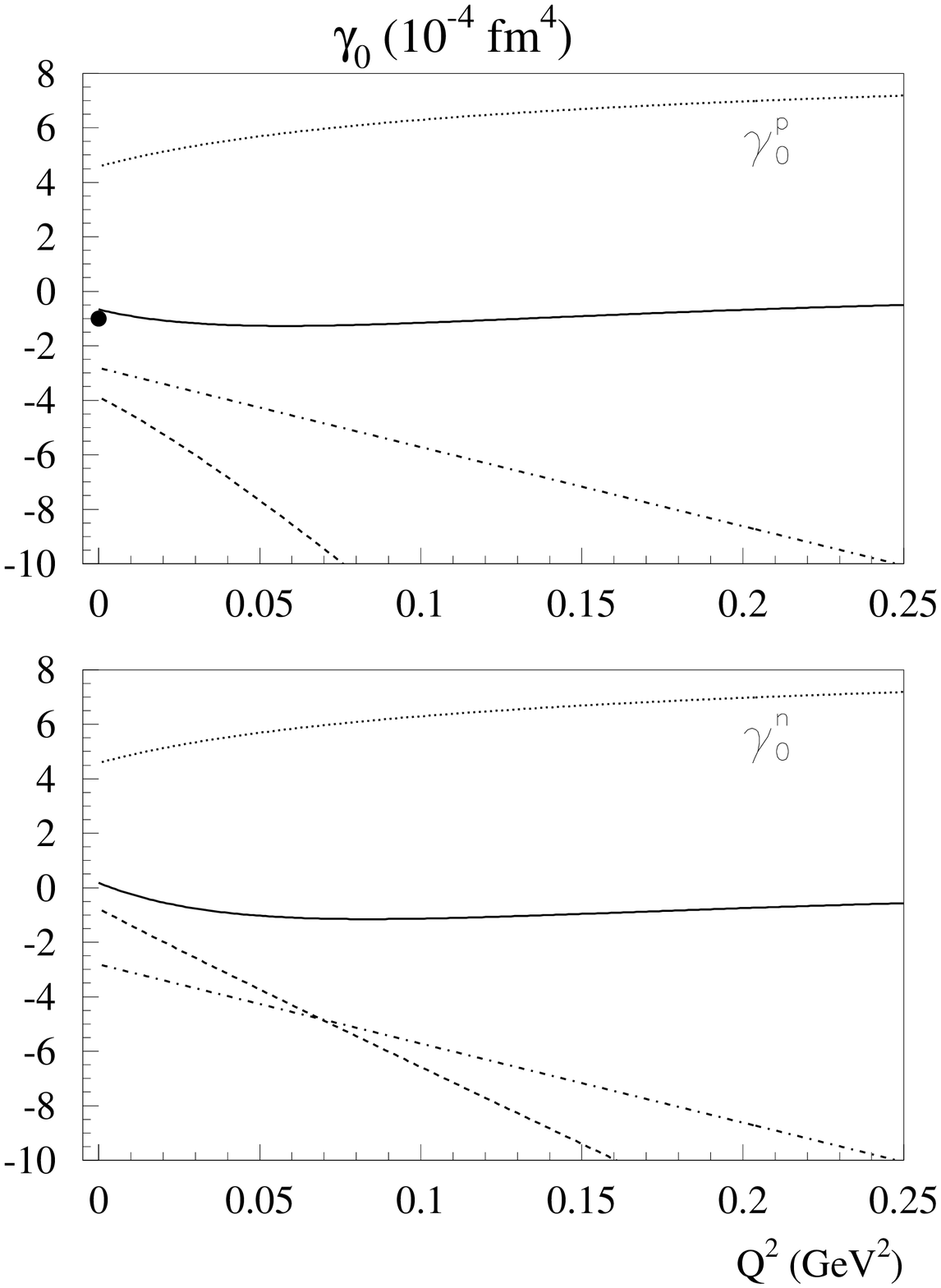}}
\vspace{0cm}
\caption{$Q^2$ dependence of the forward spin polarizability
$\gamma_0$ for proton (upper panel) and neutron (lower panel).
The solid curves represent the MAID estimate~\protect\cite{MAID,DKT01}
for the $\pi$ channel.
The dotted curves are the ${\mathcal O}(p^3)$ HBChPT results; 
the dashed curves are the ${\mathcal O}(p^3) + {\mathcal O}(p^4)$ 
HBChPT results; and the dashed-dotted curves are the 
$\Delta$ contributions evaluated in ${\mathcal O}(\varepsilon^3)$. 
The solid circle for the proton at $Q^2 = 0$ corresponds with the 
experimental value extracted from the MAMI GDH experiment 
\protect\cite{Ahr01}.}  
\label{fig:gammao_chpt}
\end{figure}

\begin{figure}[h]
\epsfysize=13cm
\centerline{\epsffile{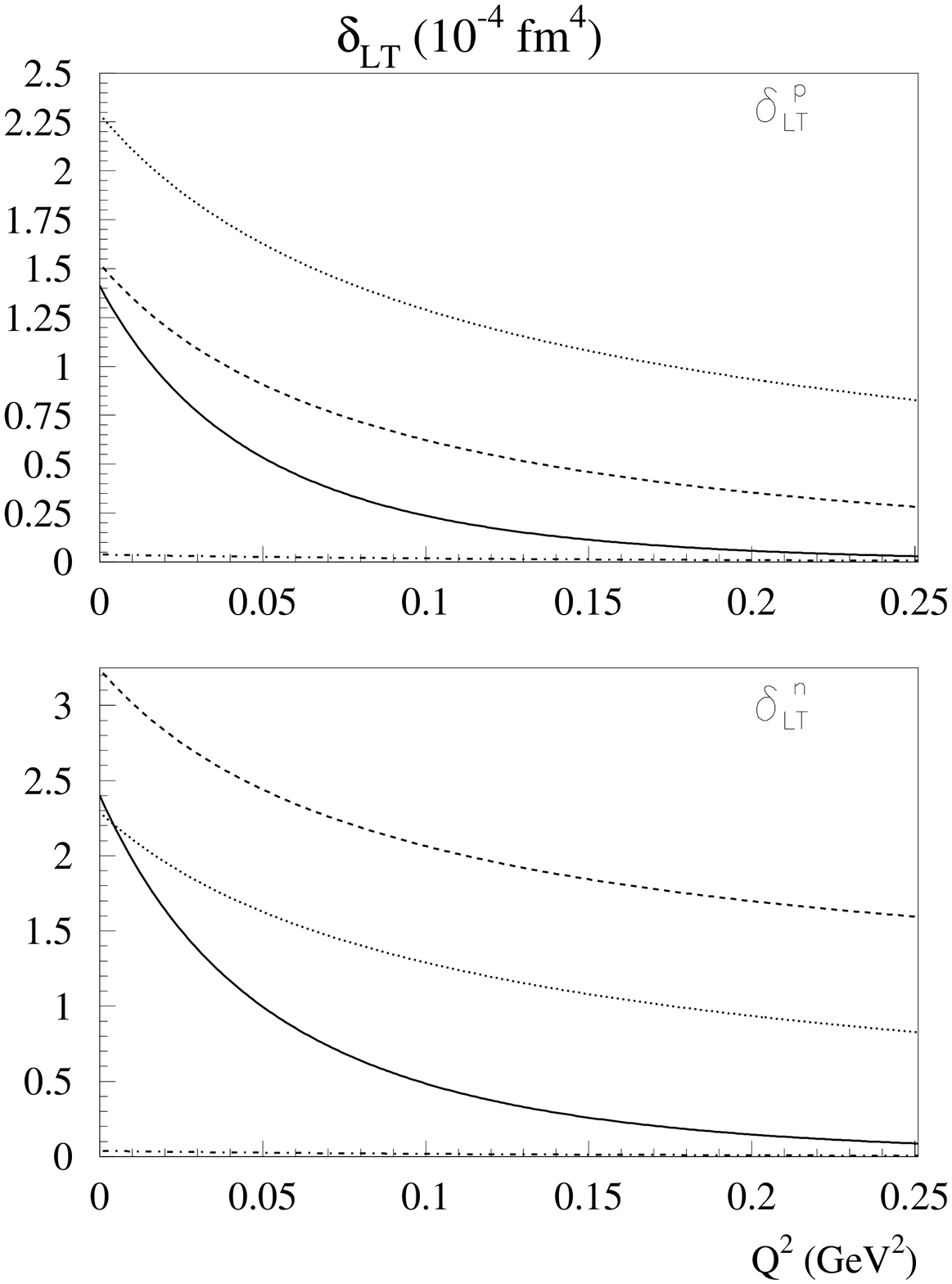}}
\vspace{0cm}
\caption{$Q^2$ dependence of the forward spin polarizability
$\delta_{LT}$ for proton (upper panel) and neutron (lower panel).
The solid curves represent the MAID estimate~\protect\cite{MAID,DKT01}
for the $\pi$ channel.
The dotted curves are the ${\mathcal O}(p^3)$ HBChPT results; 
the dashed curves are the ${\mathcal O}(p^3) + {\mathcal O}(p^4)$ 
HBChPT results; and the dashed-dotted curves (close to zero) are the 
$\Delta$ contributions evaluated in ${\mathcal O}(\varepsilon^3)$.} 
\label{fig:deltalt_chpt}
\end{figure}

The higher order terms in the LEX of Eqs.~(\ref{eq:gttlex},\ref{eq:gltlex}) 
or alternatively Eqs.~(\ref{eq:s1lex},\ref{eq:s2lex}) contain the
information on the generalized forward spin polarizabilities of the nucleon. 
In this section, we present the results for 
these generalized forward spin polarizabilities in 
HBChPT to ${\mathcal{O}}(p^4)$. Here we  include all one-particle-reducible 
diagrams but remove their pole parts as in Refs.~\cite{JKO00,Kum00}.
\newline
\indent
For the purely transverse generalized forward spin polarizability
$\gamma_0$ defined in Eq.~(\ref{eq:gttlex}), we find the expression at
 ${\mathcal{O}}(p^3)$ as~:
\begin{eqnarray}
\gamma_{0}^{{\cal O}(p^3)}(Q^2)=\frac{\alpha_{em}g_{A}^{2}}{(4\pi F_{\pi})^{2}}\cdot\frac{4}{m_{\pi}^{2}}\left[\frac{1}{3}+\frac{1}{w^2}
-\frac{4w^2+12}{3w^3\sqrt{w^2+4}}\sinh^{-1}[\frac{w}{2}]\right] \, .
\label{eq:g0p3}
\end{eqnarray}
To test the convergence of the chiral expansion, we calculate the
correction in HBChPT at ${\mathcal{O}}(p^4)$, for which we obtain the result~:
\begin{eqnarray}
\gamma_{0}^{{\cal O}(p^4)}(Q^2) = 
\frac{\alpha_{em}g_{A}^2}{(4\pi F_{\pi})^{2}M_{N}}\cdot
\frac{\pi}{1152 m_{\pi}} && 
\{-576w^2+[(-1888-448\kappa_{v})+(-416-128\kappa_{s})\tau_{3}] \nonumber \\
&&+[(366+366\kappa_{v})+(930-222\kappa_{s})\tau_{3}]\cdot\frac{1}{w^2}
\nonumber \\
&&+[(123+123\kappa_{v})+(-2787+93\kappa_{s})\tau_{3}]
\cdot\frac{1}{w}\tan^{-1}[\frac{w}{2}] \nonumber \\
&&+[(2724-732\kappa_{v})+(-1860+444\kappa_{s})\tau_{3}]
\cdot\frac{1}{w^3}\tan^{-1}[\frac{w}{2}] \nonumber \\
&&+[(272+272\kappa_{v})+(16+16\kappa_{s})\tau_{3}]\cdot\frac{1}{w^2+4} 
\nonumber \\
&&- 192\tau_{3}\cdot\frac{w^4-4w^2-24}{w^2+4}
+ 576\cdot\frac{w^4-12}{w^4+4w^2}\} \,.
\label{eq:g0p4}
\end{eqnarray}
Analogously, we find for the longitudinal-transverse generalized
forward spin polarizability $\delta_{LT}$, defined in
Eq.~(\ref{eq:gltlex}), the expressions 
for the  ${\mathcal{O}}(p^3)$ term as~:
\begin{eqnarray}
\delta_{LT}^{{\cal O}(p^3)}(Q^2)=\frac{\alpha_{em}g_{A}^{2}}{(4\pi F_{\pi})^{2}}\cdot\frac{4}{m_{\pi}^{2}}\left[
\frac{1}{3w\sqrt{w^2+4}}\sinh^{-1}[\frac{w}{2}]\right] \, ,
\label{eq:dltp3}
\end{eqnarray}
and for the ${\mathcal{O}}(p^4)$ correction as~:
\begin{eqnarray}
\delta_{LT}^{{\cal O}(p^4)}(Q^2)=
\frac{\alpha_{em}g_{A}^2}{(4\pi F_{\pi})^{2}M_{N}}\cdot\frac{\pi}{192 m_{\pi}}
&&\{(-48+24\kappa_{v})+(-24+48\kappa_{s})\tau_{3} \nonumber \\
&&+[(-162+24\kappa_{v})+(-18+24\kappa_{s})\tau_{3}]\cdot\frac{1}{w^2}
\nonumber \\
&&+[(-27-36\kappa_{v})+(-27-12\kappa_{s})\tau_{3}]\cdot
\frac{1}{w}\tan^{-1}[\frac{w}{2}] \nonumber \\
&&+[(-252-48\kappa_{v})+(36-48\kappa_{s})\tau_{3}]
\cdot\frac{1}{w^3}\tan^{-1}[\frac{w}{2}] \nonumber \\
&&+[12-(36+48\kappa_{s})\tau_{3}]\cdot\frac{1}{4+w^2}
+384\cdot\frac{3+w^2}{4w^2+w^4} \} \, .
\label{eq:dltp4}
\end{eqnarray}
\newline
\indent
At the real photon point, these forward spin polarizabilities reduce to~: 
\begin{eqnarray}
\gamma_{0}(Q^2=0)&=&\frac{\alpha_{em}g_{A}^2}{(4\pi F_{\pi})^{2}}\cdot\frac{2}{3m_{\pi}^2}\left[1-\frac{\pi m_{\pi}}{8M_{N}}[15+3\kappa_{v}+(6+\kappa_{s})\tau_{3}]\right], 
\nonumber \\
\delta_{LT}(Q^2=0)&=&\frac{\alpha_{em}g_{A}^2}{(4\pi
F_{\pi})^{2}}\cdot\frac{1}{3m_{\pi}^2}\left[1+\frac{\pi
m_{\pi}}{8M_{N}}[-3+\kappa_{v}+(-6+4\kappa_{s})\tau_{3}]\right] \, .
\end{eqnarray}
For $\gamma_0$, our expressions at the real photon point reduce 
to the result which has been
derived before in Refs.~\cite{Kum00,JiK00,Gel00}. 
Furthermore, the slopes at $Q^2 = 0$ of the generalized forward spin
polarizabilities are given by~:
\begin{eqnarray}
\left[\frac{d\gamma_{0}(Q^2)}{dQ^2}\right]_{Q^2=0}&=&\frac{\alpha_{em}g_{A}^2}{(4\pi F_{\pi})^{2}}\cdot\frac{4}{45m_{\pi}^4}\left[1
-\frac{\pi m_{\pi}}{1024M_{N}}[5451+267\kappa_{v}
+(-75+21\kappa_{s})\tau_{3}]\right], 
\nonumber \\
\left[\frac{d\delta_{LT}(Q^2)}{dQ^2}\right]_{Q^2=0}&=&\frac{\alpha_{em}g_{A}^2}{(4\pi F_{\pi})^{2}}\cdot\frac{-1}{18m_{\pi}^4}\left[1+
\frac{\pi m_{\pi}}{80M_{N}}[54-9\kappa_{v}-(27+24\kappa_{s})\tau_{3}]\right] 
\, .
\end{eqnarray}
\newline
\indent
Besides the $\pi N$ loop contribution, we also estimated the effects
of the $\Delta$ contributions to the forward spin
polarizabilities in the small scale expansion to order 
${\mathcal{O}}(\varepsilon^3)$, where $\varepsilon$ stands for a soft
momentum, the pion mass or the mass difference between $\Delta$ and
nucleon, denoted by $\Delta=M_{\Delta}-M_{N}$.
The $\Delta$ effects enter through both a $\Delta$-pole contribution  
as well as $\pi \Delta$ loop contributions (see Fig.~\ref{fig:delta}).
The leading order $\Delta$ contributions to the 
generalized forward spin polarizabilities are given as~:
\begin{eqnarray}
\gamma_{0}^{\Delta}(Q^2)&=&\frac{-\alpha_{em}}{9}(\frac{G_{1}}{M_{N}})^{2}\cdot\frac{\Delta^{2}+Q^{2}}{\Delta^{4}} \nonumber \\
&-&\frac{32\alpha_{em}}{27}\frac{g_{\pi\Delta N}^{2}}{(4\pi F_{\pi})^{2}}
\int^{1}_{0}dx \;\frac{x^3}{m_{0}^{2}}(\mu_{0}^{2}-1)^{-2}\left[\mu_{0}^{2}
+2-3\mu_{0}\frac{\ln[\mu_{0}+\sqrt{\mu_{0}^{2}-1}]}{\sqrt{\mu_{0}^{2}-1}}
\right] \nonumber \\
&-&\frac{8\alpha_{em}}{27}\frac{g_{\pi\Delta N}^{2}}{(4\pi F_{\pi})^{2}}
\int^{1}_{0}dx \; \frac{x^4(1-2x)Q^{2}}{m_{0}^{4}}(\mu_{0}^{2}-1)^{-3}
\nonumber \\
&&\hspace{4cm}\times\left[11\mu_{0}^{2}+4-(6\mu_{0}^{3}+9\mu_{0})
\frac{\ln[\mu_{0}+\sqrt{\mu_{0}^{2}-1}]}{\sqrt{\mu_{0}^{2}-1}}\right]\, , 
\label{eq:poldelta} 
\end{eqnarray}
and 
\begin{eqnarray}
\delta_{LT}^{\Delta}(Q^2)&=&
\frac{-32\alpha_{em}}{27}\frac{g_{\pi\Delta N}^{2}}{(4\pi F_{\pi})^{2}}
\int^{1}_{0}dx \;\frac{x^3}{m_{0}^{2}}(\mu_{0}^{2}-1)^{-2}\left[\mu_{0}^{2}
+2-3\mu_{0}\frac{\ln[\mu_{0}+\sqrt{\mu_{0}^{2}-1}]}{\sqrt{\mu_{0}^{2}-1}}
\right] \nonumber \\
&+&\frac{16\alpha_{em}}{9}\frac{g_{\pi\Delta N}^{2}}{(4\pi F_{\pi})^{2}}
\int^{1}_{0}dx \;\frac{x^{2}(1-2x)}{m_{0}^{2}}(\mu_{0}^{2}-1)^{-1}
\left[1-\mu_{0}\frac{\ln[\mu_{0}+\sqrt{\mu_{0}^{2}-1}]}{\sqrt{\mu_{0}^{2}-1}}
\right]\, , 
\label{eq:pol2delta}
\end{eqnarray}
with $m_{0}\equiv \sqrt{m_{\pi}^{2}+x(1-x)Q^2}$ and $\mu_{0}\equiv \frac{\Delta}{m_{0}}$.
$G_{1}$ and $g_{\pi\Delta N}$ are the leading order $\gamma \Delta N$ and $\pi\Delta N$ coupling constants respectively.
In the large $N_{C}$ limit of QCD, they are related 
with $\kappa_{v}$ and $g_{A}$ as~:
\begin{equation}
G_{1}=\frac{3}{2\sqrt{2}}\kappa_{v},\,\,\,\,\,\,g_{\pi\Delta N}=\frac{3}{2\sqrt{2}}g_{A}.
\end{equation}
Note that Eqs.~(\ref{eq:poldelta}) and (\ref{eq:pol2delta}) apply when 
$Q^2 < 4 (\Delta^2 - m_\pi^2) \simeq 0.268$~GeV$^2$. Analogous
formulas can be found for the range $Q^2 > 4 (\Delta^2 - m_\pi^2)$,
which we don't specify here as they fall outside 
the range which we show in the following figures.  
\newline
\indent
At the real photon point, the expressions for the $\Delta$
contributions reduce to~:
\begin{eqnarray}
\gamma^{\Delta}_{0}(Q^2=0) \,&=&\,
-\frac{\alpha_{em}}{9}(\frac{G_{1}}{M_{N}})^{2}\cdot\frac{1}{\Delta^{2}}
\nonumber \\
&+&\frac{\alpha_{em}g_{\pi\Delta N}^{2}}{(4\pi F_{\pi})^{2}}
\cdot\frac{8}{27m_{\pi}^{2}}(\mu^{2}-1)^{-2}
\left[-\mu^2-2+3\mu\frac{\ln[\mu+\sqrt{\mu^2-1}]}{\sqrt{\mu^2-1}}\right]\, ,
\nonumber \\
\delta_{LT}^{\Delta}(Q^2=0)\,&=&\,
\frac{\alpha_{em}g_{\pi\Delta N}^{2}}{(4\pi F_{\pi})^{2}}
\cdot\frac{8}{27m_{\pi}^{2}}(\mu^{2}-1)^{-2}\left[-2\mu^2-1+(\mu^3+2\mu) 
\frac{\ln[\mu+\sqrt{\mu^2-1}]}{\sqrt{\mu^2-1}}\right] \, ,
\end{eqnarray}
with $\mu=\frac{\Delta}{m_{\pi}}$. Furthermore, the slopes at $Q^2 = 0$ 
of the generalized forward spin polarizabilities are given by~:
\begin{eqnarray}
\left[\frac{d\gamma_{0}^{\Delta}(Q^2)}{dQ^{2}}\right]_{Q^{2}=0}&=&
-\frac{\alpha_{em}}{9}(\frac{G_{1}}{M_{N}})^{2}\cdot\frac{1}{\Delta^{4}}
\nonumber \\
&&\hspace{-1cm}+\frac{\alpha_{em}g_{\pi\Delta N}^{2}}{(4\pi F_{\pi})^{2}}
\cdot\frac{8}{405m_{\pi}^{4}}
\left\{ (\mu^{2}-1)^{-2}
\left[4\mu^2+4-9\mu\frac{\ln[\mu+\sqrt{\mu^2-1}]}{\sqrt{\mu^2-1}}
\right] \right. \nonumber \\
&&\hspace{2.65cm} + \,(\mu^{2}-1)^{-3}
\left[-4\mu^4+11\mu^{2}+8 \right. \nonumber \\
&& \left. \left. \hspace{4.65cm}+(3\mu^{3}-18 \mu)
\frac{\ln[\mu+\sqrt{\mu^2-1}]}{\sqrt{\mu^2-1}}\right] \right\} \, ,
\nonumber \\
\left[\frac{d\delta_{LT}^{\Delta}(Q^2)}{dQ^{2}}\right]_{Q^{2}=0}&=&\frac{\alpha_{em}g_{\pi\Delta N}^{2}}{(4\pi F_{\pi})^{2}}
\cdot\frac{2}{405m_{\pi}^{4}} \nonumber \\
&\times& \left\{(\mu^{2}-1)^{-1}
\left[6-9\mu\frac{\ln[\mu+\sqrt{\mu^2-1}]}{\sqrt{\mu^2-1}}\right]
\right. \nonumber \\
&& +\,(\mu^{2}-1)^{-2}\left[7\mu^{2}+16+(9\mu^{3}-36\mu)\frac{\ln[\mu+\sqrt{\mu^2-1}]}{\sqrt{\mu^2-1}}\right]
\nonumber \\
&& \left. +\,(\mu^{2}-1)^{-3}\left[-16\mu^{4}-44\mu^{2}+60\mu^{3}
\frac{\ln[\mu+\sqrt{\mu^2-1}]}{\sqrt{\mu^2-1}}\right] \right\} \, .
\end{eqnarray}
\indent
In Fig.~\ref{fig:gammao_chpt}, we compare the HBChPT results for
$\gamma_0(Q^2)$ with the phenomenological estimates using the MAID model. 
For $\gamma_0$, the experimental value at the real photon point has
been obtained by the GDH experiment \cite{Ahr01} and is indicated.
The MAID estimate at $Q^2 = 0$ is in good agreement with the
experimental value. 
By comparing the HBChPT results for $\gamma_0$ at  
${\mathcal{O}}(p^3)$ and at ${\mathcal{O}}(p^4)$, one sees that 
for both proton and neutron, the  ${\mathcal{O}}(p^4)$ correction term
is of opposite sign as the leading order term, and its magnitude is
even larger than the leading order term, as has been noted before in 
Refs.~\cite{Kum00,JiK00}. 
By comparing the HBChPT result with the estimate as  
obtained with MAID, one sees that there is a large cancellation
involved in $\gamma_0$ over the whole $Q^2$ region. 
Although the ${\mathcal{O}}(p^3)$ and ${\mathcal{O}}(p^4)$ terms are
of opposite sign, and also display a large cancellation, 
their large magnitudes indicate that no convergence has been reached 
for $\gamma_0$ at ${\mathcal{O}}(p^4)$. 
One may try to improve the HBChPT calculation by adding explicit
$\Delta$ degrees of freedom. To this end, we also show 
in Fig.~\ref{fig:gammao_chpt} the 
$Q^2$ dependence of the $\Delta$ contribution to $\gamma_0$ 
calculated in the small scale expansion to order 
${\mathcal{O}}(\varepsilon^3)$ according to Eq.~(\ref{eq:poldelta}).
One sees from  Fig.~\ref{fig:gammao_chpt} that the
${\mathcal{O}}(\varepsilon^3)$ correction term 
is sizeable and reduces the 
${\mathcal{O}}(p^3)$ result to come closer to the phenomenological
result. However, quantitatively the result at order 
${\mathcal{O}}(\varepsilon^3)$ is still not satisfying 
when compared to the phenomenological result, and
suggests that one needs to estimate the terms of the order 
${\mathcal{O}}(\varepsilon^4)$ in a future work.
\newline
\indent
The corresponding HBChPT results for the forward spin polarizability 
$\delta_{LT}(Q^2)$ are shown in Fig.~\ref{fig:deltalt_chpt}. 
In contrast to $\gamma_0$, one sees that $\delta_{LT}$ receives a much
smaller correction at ${\mathcal{O}}(p^4)$. 
For the proton, the ${\mathcal{O}}(p^4)$ correction term 
reduces the ${\mathcal{O}}(p^3)$ result. By comparing the HBChPT result 
at ${\mathcal{O}}(p^4)$ with the phenomenological estimate, one sees a
reasonable agreement for the proton. For the neutron however, the correction
at  ${\mathcal{O}}(p^4)$ increases the  ${\mathcal{O}}(p^3)$ result and lies
about  20~\% above the value predicted by MAID. 
This discrepancy may however again be due to the uncertainty in the
phenomenological estimate of $\sigma_{LT}$ for the neutron, 
which enters the {\it rhs} of Eq.~(\ref{eq:deltalt}), 
as was also seen for the neutron $I_2$ result in Fig.~\ref{fig:i2}. 
Furthermore, we see that for $\delta_{LT}$ the correction due to 
explicit $\Delta$-resonance degrees of freedom is very small, making the 
forward spin polarizability 
$\delta_{LT}$ a very useful observable to study the transition from
the low $Q^2$ regime to the perturbative regime at large $Q^2$. 
\newline
\indent
Recently, doubly polarized asymmetries for inclusive electron
nucleon scattering were measured at Jefferson Lab (JLab) using longitudinally
polarized electrons incident on a longitudinally polarized nucleon 
\cite{CLAS,HallA}. These experiments are mainly sensitive to the
nucleon structure function $g_1$, with some small admixture of $g_2$. 
For the proton, the CLAS Collaboration \cite{CLAS} measured the first moment of
$g_1$ down to $Q^2 \simeq 0.15$~GeV$^2$. For the neutron, 
the Hall A Collaboration \cite{HallA} measured the first moment of $g_1^n$ 
down to $Q^2 \simeq 0.10$~GeV$^2$, while a forthcoming Hall A experiment
will extend this down to  $Q^2 \simeq 0.02$~GeV$^2$ \cite{HallAb}. 
These measurements map out the transition from the GDH sum rule value
at $Q^2 = 0$ to the DIS value at large $Q^2$.
In this way, they allow us to study quantitatively the transition from
the resonance dominated regime at low $Q^2$ to the partonic 
regime at large $Q^2$.  
In terms of the virtual photon absorption cross sections $\sigma_{TT}$
and $\sigma_{LT}$ of Eqs.~(\ref{eq:stt},\ref{eq:slt}), the JLab
experiments \cite{CLAS,HallA} 
with a longitudinally polarized target, are mainly sensitive to $\sigma_{TT}$.
Therefore, it will be interesting to extract from these data 
the $Q^2$ dependence of $\gamma_0(Q^2)$ for 
the proton and neutron according to Eq.~(\ref{eq:gammao}). 
Compared to the first moment of $g_1$, the forward spin polarizability 
involves the third moment of $g_1$ (see Eq.~(\ref{eq:gammao})). 
Therefore, this observable receives only a negligeable small 
contribution from the unmeasured region at small-$x$, and can be
directly extracted from the $x$-range as accessed by the present
experiments at the lower $Q^2$.
The extraction of $\delta_{LT}(Q^2)$ requires the measurement of 
$\sigma_{LT}$. This can be achieved by measuring the 
inclusive electroabsorption cross section for longitudinally
polarized electrons on a transversely polarized nucleon target. 
Such measurements, giving access to the spin structure function $g_2$,  
are presently underway at JLab 
both in Hall A \cite{Averett} and Hall C \cite{Rondon}.

\section{Conclusions}

In conclusion, in this paper we 
studied spin-dependent sum rules for forward virtual Compton scattering
(VVCS) off the nucleon in HBChPT at order ${\mathcal{O}}(p^4)$. 
We showed in particular that the 
Burkhardt-Cottingham sum rule is satisfied in HBChPT to this order. 
We have also noticed that the ${\mathcal{O}}(p^4)$ prediction remains
close to the phenomenological sum rule evaluation, in the range up to
$Q^2 \simeq 0.25$~GeV$^2$. This relatively wide range indicates that the first
moment of $g_2$ is a promising observable to bridge the gap between
the HBChPT description at the lower $Q^2$ and the perturbative QCD
result at the larger $Q^2$.
\newline
\indent
Furthermore, we related the higher terms in the
low energy expansion of the VVCS amplitude, to generalized forward spin
polarizabilities of the nucleon.
We then calculated these generalized forward spin
polarizabilities of the nucleon at ${\cal O}(p^4)$ in HBChPT.
The result for $\gamma_{0}(Q^2)$ indicates a large ${\cal O}(p^4)$ 
correction to the leading order result. 
The corresponding result for $\delta_{LT}(Q^2)$ however 
shows that the convergence of the heavy
baryon and chiral expansion is better behaved, and 
receives only a negligeable contribution from $\Delta$ degrees of freedom.
It will be important to extract the information on the generalized 
forward spin polarizabilities from experiments at different $Q^2$,
such as are presently performed at JLab. This will allow to quantify the
transition from the resonance dominated regime at $Q^2 = 0$ to the
partonic regime at large $Q^2$.
For $\delta_{LT}(Q^2)$, 
our result also indicates that measuring the value of the integral related 
to $\delta_{LT}$ at $Q^2 = 0.02$~GeV$^2$, as accessible in the
experiment of Ref.~\cite{HallAb},  
will be enough to extrapolate to the value at $Q^2=0$.

\section*{Acknowledgements}

This work was supported by the Deutsche Forschungsgemeinschaft (SFB443). 
The authors also like to thank for the support and hospitality 
of the ECT* (Trento) at the Collaboration meeting 
``Baryon structure probed with quasistatic electromagnetic
fields'' in 2002, during which this work was discussed.  
Furthermore, the authors like to thank 
D. Drechsel for his interest in this work and for numerous discussions.


\end{document}